\documentclass[9pt,twocolumn,twoside]{pnas-new}

\templatetype{pnasinvited} 

\usepackage{graphicx} 
\usepackage[round]{natbib}
\usepackage{amsmath}
\usepackage{amssymb}

\makeatletter
\AtBeginDocument{%
  \fancyhead[LO]{}%
  \fancyhead[RO]{}%
  \fancyhead[LE]{}%
  \fancyhead[RE]{}%

  \fancypagestyle{plain}{%
    \fancyhead{}
    \fancyfoot[R]{\footerfont PNAS\hspace{7pt}|\hspace{7pt}\textbf{\today}\hspace{7pt}|\hspace{7pt}vol. XXX\hspace{7pt}|\hspace{7pt}no. XX\hspace{7pt}|\hspace{7pt}\textbf{\thepage\space of\space\pageref{LastPage}}}%
    \fancyfoot[L]{\footerfont\@doi}%
  }%

  \pagestyle{plain}%
}
\makeatother

\begin{document}

\title{Exploring Exoplanet Dynamics with JWST: \\ Tides, Rotation, Rings, and Moons}


\author[a, b, 1]{Sarah C. Millholland}
\author[c]{Joshua N.\ Winn}

\affil[a]{Department of Physics, Massachusetts Institute of Technology, Cambridge, MA 02139, USA}
\affil[b]{Kavli Institute for Astrophysics and Space Research, Massachusetts Institute of Technology, Cambridge, MA 02139, USA}
\affil[c]{Department of Astrophysical Sciences, Princeton University, Princeton, NJ 08544, USA}

\leadauthor{Millholland \& Winn}


\authorcontributions{The authors contributed equally to this work.}
\authordeclaration{The authors declare no competing interests.}
\correspondingauthor{\textsuperscript{1}To whom correspondence should be addressed. E-mail: sarah.millholland@mit.edu}


\begin{abstract}

Although nearly 6{,}000 exoplanets are currently known, in most cases our knowledge is limited to a handful of the planet's orbital characteristics and bulk properties such as radius and mass. The James Webb Space Telescope (JWST) can expand our knowledge not only by probing exoplanet atmospheres, but also by measuring additional orbital and physical properties of exoplanets, thanks to its superior light-gathering power and measurement precision. Here, we describe the potential of JWST to unveil dynamical phenomena that were previously beyond our reach, such as tidal distortion and inflation, rotational flattening, planetary rings, and moons.
\end{abstract}

\dates{This manuscript was compiled on \today}
\doi{\url{www.pnas.org/cgi/doi/10.1073/pnas.XXXXXXXXXX}}

\maketitle
\thispagestyle{firststyle}
\ifthenelse{\boolean{shortarticle}}{\ifthenelse{\boolean{singlecolumn}}{\abscontentformatted}{\abscontent}}{}




Spectroscopy is fundamental to astronomy, and
the launch of JWST
in December 2021 was a giant leap for astronomical spectroscopy.
Approximately
75\% of all JWST observing programs are devoted to spectroscopy,\footnote{https://www.stsci.edu/jwst/science-execution/approved-programs/general-observers} including
many efforts to study the atmospheres of extrasolar planets. JWST has already transformed the field of exoplanetary atmospheres by providing major improvements in wavelength coverage, photometric precision, and stability over
previous datasets \citep[e.g.][]{2023Natur.614..649J, 2023Natur.614..659R, 2023ApJ...956L..13M}.

A separate topic within exoplanetary science
that JWST has not yet advanced significantly is {\it dynamics}, a term that broadly refers to the motion, interactions, and evolution of planets. For the purpose of this article,
we focus on a limited set of subjects involving the gravitational interactions between planets, the perturbations of their orbits 
and shapes due to rotation and tides, 
and the possibilities of moons and rings analogous to those
of Jupiter and Saturn.
Studying these phenomena does not always require
spectroscopy.
A small fraction of planets have
orbital planes that happen to be
nearly parallel to our line of sight, causing the planet
to transit directly in front of the star and block a portion of its radiation.
These planets are characterized using {\it time-series photometry},
the measurement of total brightness versus time (or {\it light curve}).

The most precise light curves are obtained
with space telescopes because in space, a telescope is free of 
the corrupting effects of Earth's fluctuating atmosphere.
But even in space, there is an ultimate limit to the precision of a light curve arising from the random fluctuations of photons.
The maximum signal-to-noise ratio is the square root 
of the expected number of photons detected during the time interval of the
measurement. Thus, the precision can be improved by increasing the light-collecting area.

In this respect, JWST's main advantage over its predecessors is its light-collecting area of 25~m$^2$, much larger than the Hubble Space Telescope's
4~m$^2$ and the Spitzer Space Telescope's
0.6~m$^2$.
Moreover, JWST's instruments introduce fewer artifacts and allow a closer approach to the fundamental photon-noise limit \citep[e.g][]{2022A&A...661A..80J}.
JWST's location is also
superb. Millions of kilometers away from the Earth, the telescope enjoys dark and stable conditions
where it can stare continuously for long
stretches of time. For example, a JWST light curve
of the ``sub-Neptune'' GJ\,1214\,b spans 48 hr and has a precision equivalent to $\approx$100 parts per million (ppm) per 1-minute sample, within 10\% of the photon-noise limit \citep{2023Natur.620...67K}. We will use this level of precision as a benchmark throughout this paper, while noting that it is only achievable for the brightest stars.

In this article, we discuss some dynamics-related effects that may be observable with JWST: tidal orbital decay, tidal distortion of a planet's shape, tidal heating and inflation of low-density planets,
rotational oblateness and obliquity, and
planetary rings and moons. These topics have not yet been thoroughly explored by JWST, owing to the community's focus on atmospheric characterization and the necessity in some applications to target long-period planets, the most favorable of which may yet to be detected (as described below). Exoplanet dynamics is a broad field encompassing many subjects, but our scope
is limited mainly to areas where JWST can make substantial advances -- mainly with time-series photometry -- including first-ever detections of phenomena. Moreover, we provide only order-of-magnitude estimates of these effects rather than the detailed modeling that will be necessary for planning realistic programs.
Before delving into these topics, we first remark on our current knowledge of the potential targets for these observations.

\section*{The best and brightest stars}

The success or failure of each type of observation described below
depends not only on the characteristics of the telescope, but also on knowing where to point it.
Subtle signals can only be detected
when the host star is very bright and the photon-noise limit is correspondingly low. 
Fortunately, the advent of JWST was preceded by several major surveys for transiting planets.

\begin{figure}[th!]
\centering
\includegraphics[width=\linewidth]{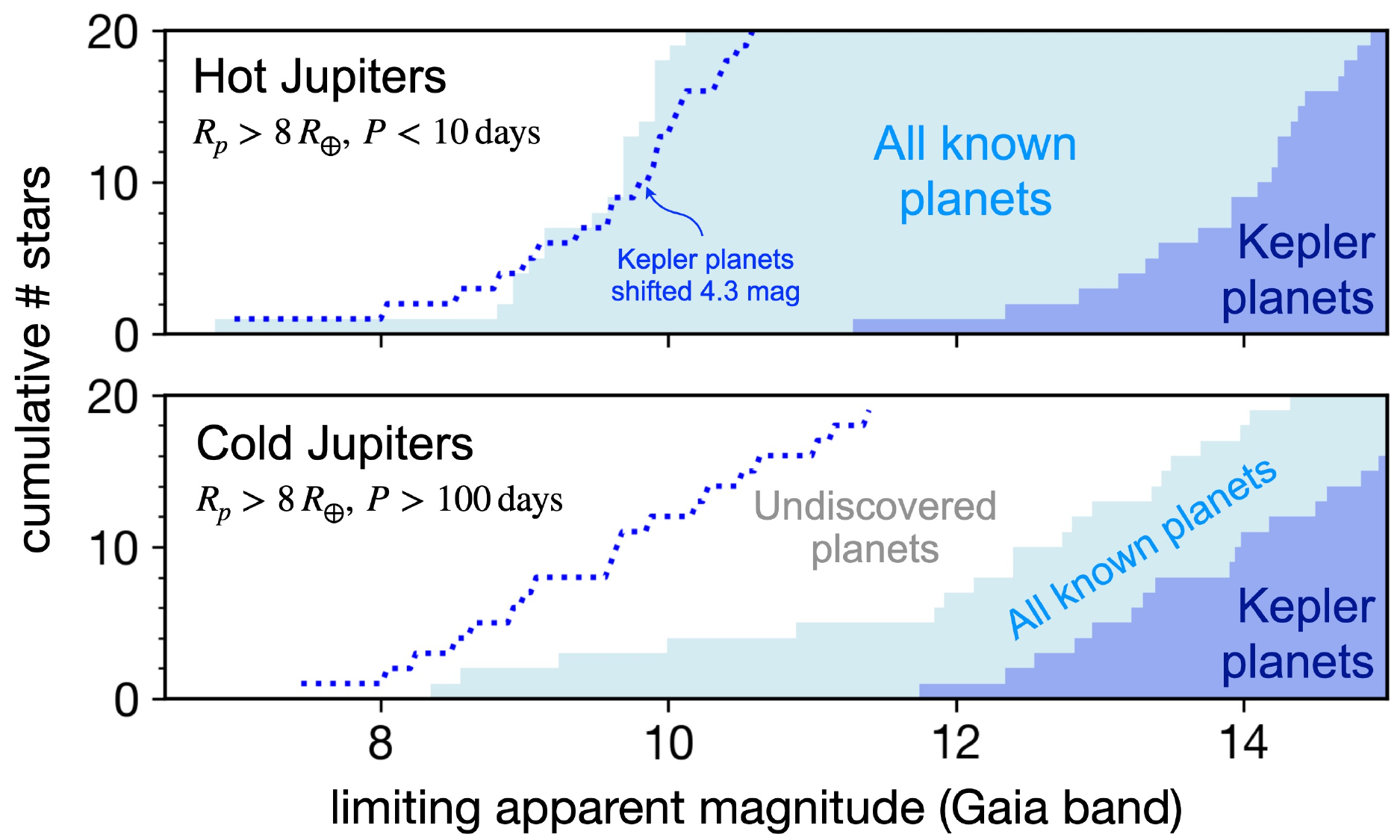}
\caption{{\bf The brightest stars with transiting planets.}
The histograms are cumulative distributions of the apparent Gaia magnitudes of stars with known transiting planets, for the Kepler survey (dark blue) and over the entire sky (light blue).
One would expect the distribution for a complete all-sky
catalog to resemble that of the Kepler survey
shifted leftward by 4.3~mag (dotted blue line).
For hot Jupiters (top panel), this is nearly
the case, suggesting that many of the brightest hosts are already known.
For planets with periods of months or longer,
such as cold Jupiters (bottom panel), many brighter hosts surely exist but remain undiscovered.
Because it is based on the Kepler survey, this comparison is restricted to Sun-like stars with effective temperatures 4000--6000~K and does not address the completeness of surveys of lower-mass stars that emit most of their radiation at infrared wavelengths.}
\label{fig:kepler_vs_allsky}
\end{figure}

From 2009 to 2018,
NASA's Kepler space telescope \citep{2010Sci...327..977B} discovered thousands of transiting planets,
including hundreds of multiple-planet systems
that have proven fruitful for dynamical studies \citep[e.g.][]{2014ApJ...790..146F, 2023ASPC..534..863W}.
Kepler was sensitive to planets over a wider range of radii and orbital periods than any other survey to date. However, Kepler surveyed only the stars within a relatively
small field of view, leaving undiscovered the analogous systems around brighter stars that are spread over the rest of the sky.
Since 2018, NASA's Transiting Exoplanet Survey Satellite (TESS)
has been surveying the entire sky \citep{2015JATIS...1a4003R}.
If planetary systems are spread randomly in all directions,
then to find the
same number of systems as the Kepler
survey, an all-sky survey would only need to
look one-seventh as far away, at stars that are brighter
by 4.3~magnitudes or nearly a factor of 50 in flux.\footnote{The
ratio of limiting distances is
$(4\pi/\Omega)^{1/3}$, where $\Omega= 115$~square degrees
is the field of view of the four-year ``Kepler prime'' survey.}
Thus, if all-sky surveys were as complete as the Kepler survey, the distribution of stellar brightnesses of known systems would resemble that of the Kepler sample after shifting it by 4.3~mag. We illustrate this with examples of two planet populations.
The top panel of Figure \ref{fig:kepler_vs_allsky} shows that this is indeed the case for hot Jupiter hosts, suggesting that the most favorable such systems
have already been discovered.
The bottom panel shows that much more room for improvement is available for 
giant planets in wider orbits, which are needed to detect many of the physical effects described below. Those attempting such detections should bear in mind that the best targets are yet to be found, perhaps by extensions
of the TESS mission or
the forthcoming PLATO mission~\citep{2014ExA....38..249R}.

The advantage of the brightest stars is a lower noise level, in principle, although JWST observations need to be planned carefully to avoid saturating the detectors. The telescope's four instruments have numerous modes differing in wavelength coverage and spectral resolution. The aforementioned observation of GJ~1214 was conducted with the Mid-Infrared Instrument (MIRI), the only option for wavelengths longer than 5~$\mu$m. For many applications described below, the exact wavelength coverage or resolution is not as important as maintaining linearity while maximizing the rate of recorded photons in a broad bandpass. The Single Object Slitless Spectroscopy (SOSS) mode of the Near-Infrared Imager and Slitless Spectrograph (NIRISS) was specifically designed for time-series observations of bright stars, resulting in some of the lowest reported noise levels in the literature (e.g., $\approx$100~ppm per 1-minute sample for WASP-39 \cite{Carter+2024}). The most popular instrument for transit observations is the Near Infrared Spectrograph (NIRSpec), which features an especially wide aperture and fast readout mode to capitalize on the brightest stars \citep{Birkmann+2022}. Observers using both of these modes are developing algorithms to mitigate the effects of time-correlated noise due to readout electronics and variations in the telescope pointing.

\section*{Orbital decay}

We begin our exploration of dynamical phenomena with tidal orbital decay, a subject that has been discussed for nearly 30 years. 
One of the earliest theoretical papers about hot Jupiters showed that they are formally unstable to tidal orbital decay \citep{1996ApJ...470.1187R}. Even when there is no tidal dissipation within the planet due to a circular orbit and a synchronous and aligned spin, the planet's gravity raises a tidal bulge on the star and excites waves within its interior. The friction associated with tidal oscillations gradually drains energy from the system
and causes the planet to spiral inward \citep{1980A&A....92..167H}.
For one planet, WASP-12\,b, the interval between transits has been observed to decrease by 30 milliseconds per year, at which rate the planet will be destroyed within a few million years \citep{2016A&A...588L...6M, 2020ApJ...888L...5Y}. 

It would be valuable to detect orbital decay in more systems, since the rates depend sensitively on the
poorly understood mechanisms by which tidal oscillations are
excited and dissipated \citep{2024PSJ.....5..163A, 2025arXiv250108992M}. In a simplified model for these oscillations, the
decay timescale is \citep{2009ApJ...698.1357J, 2010A&A...516A..64L}
\begin{equation}
\tau_{\mathrm{decay}} \equiv \left\lvert\frac{a}{\dot{a}}\right\rvert = P \left(\frac{Q_{\star}'}{9\pi}\right)\left(\frac{M_{\star}}{M_p}\right)\left(\frac{a}{R_{\star}}\right)^{\!\!5},
\label{eq: tau_a}
\end{equation}
where $P$ is the planet's orbital period, $M_{\star}$ and $R_{\star}$ are the stellar mass and radius, 
$M_p$ is the planet mass, $a$ is the semi-major axis, and $Q_{\star}'$ is the reduced tidal quality factor that 
quantifies the rate of tidal dissipation \citep{1966Icar....5..375G, 2004ApJ...610..477O}.
The topmost bar in Figure \ref{fig: mechanism regimes} illustrates the decay regimes for a Jupiter-like planet around a Sun-like star with $Q_{\star}'=10^5$. Giant planets with periods shorter than a day are expected to decay on timescales of $\tau_{\mathrm{decay}}\lesssim10$ Myr \citep[e.g.][]{2020ApJ...888L...5Y, 2024PSJ.....5..163A}. (Note $\tau_{\mathrm{decay}}=10$ Myr corresponds to $\dot{P}\approx-10$ ms/yr for a planet on a 1 day orbit around a solar-mass star).

Observing a sequence of transit times allows a planet's orbital period to be determined precisely. The timing precision is \citep{2008ApJ...689..499C}
\begin{equation}
\sigma_{t_c} \approx 1\,{\rm sec}
\left( \frac{\sigma_1}{100\,{\rm ppm}} \right)
\left( \frac{\delta}{0.01} \right)^{-1}
\left( \frac{\tau_{\rm ing}}{10\,{\rm min}} \right)^{1/2},
\end{equation}
where $\sigma_1$ is the photometric noise level in a 1-minute sample,
$\tau_{\rm ing}$ is the duration of the crucial ``ingress''
phase of the transit (when the planet's silhouette moves across
the rim of the stellar disk),
and $\delta$ is fraction of light
blocked by the planet.
The numerical
values in this scaling relation
were chosen to represent a nominal JWST observation
of a hot Jupiter.
In the course of its sky survey,
TESS observes each hot Jupiter every few years
for about a month at a time, but since $\sigma_1$ scales inversely with telescope diameter, each JWST transit time is approximately 60 times more precise than a TESS transit time. Thus, JWST observations
of the closest-orbiting hot Jupiters will increase sensitivity
to period changes. Such observations are being planned for the purpose of studying
the atmospheres of ultra-hot Jupiters, and
orbital decay might be detected as a byproduct of these
investigations, as was done recently for WASP-43\,b \citep{2025A&A...694A.233B}.

\section*{Tidal distortion}
\label{sec: tidal distortion}

\begin{figure*}
\centering
\includegraphics[width=0.9\linewidth]{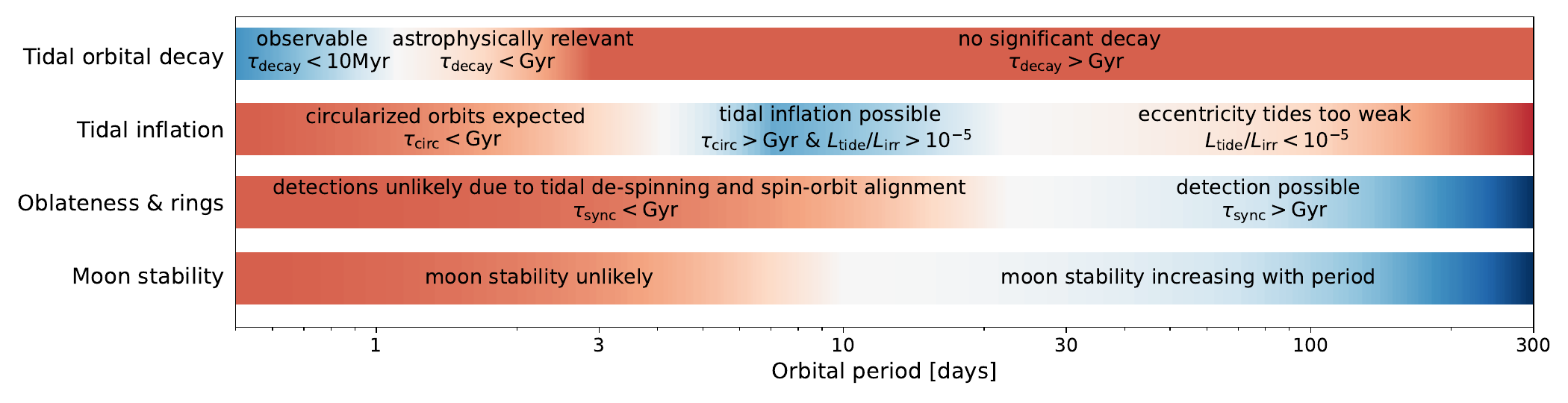}
\caption{\textbf{Detectability of various phenomena as a function of orbital period.} We consider a fiducial system consisting of a Jupiter-like planet with $Q'=10^6$ orbiting a Sun-like star with $Q_{\star}' = 10^7$. {\bf Tidal orbital decay} is fast ($\tau_{\mathrm{decay}} < 10$\,Myr, equation \ref{eq: tau_a}) for very short-period planets and is observable as a gradual reduction in the time between transits. For planets with longer periods, the changing transit period is not detectable on human timescales, but is still astrophysically relevant if $\tau_{\mathrm{decay}} <$\,Gyr. {\bf Tidal inflation} requires the planet's orbit to be eccentric and close enough to the star for a significant tidal luminosity $L_{\mathrm{tide}}$ relative to the incident stellar power $L_{\mathrm{irr}}$. We used expressions from \cite{2010A&A...516A..64L} for $L_{\mathrm{tide}}$ and the tidal circularization timescale $\tau_{\mathrm{circ}}$. In the eccentric regime, we assumed a fiducial value, $e = 0.2$. {\bf Oblateness and rings} are only expected to be detectable for planets in wide enough orbits to avoid tidal spin-down. We used the expression for the tidal synchronization timescale $\tau_{\mathrm{sync}}$ from \cite{1996Icar..122..166G}, assuming an initial rotation rate equal to Jupiter's current rate.
{\bf Moons} are dynamically unstable for planets with very short orbital periods due to the planet's small Hill radius and the effects of tidal evolution. The approximate moon instability cutoff of $P\approx10$\,days is based on \cite{2021PASP..133i4401D}. }
\label{fig: mechanism regimes}
\end{figure*}

The strong tidal forces between stars and short-period planets have other consequences. Eccentric orbits, nonsynchronous spins, or tilted spin-axes cause the planet's tidal bulge to vary in time, leading to internal heat dissipation  \citep[e.g.][]{1880RSPT..171..713D,  1966Icar....5..375G,  2008ApJ...678.1396J, 2008ApJ...681.1631J}. 
Dissipation causes the orbits of close-in planets to circularize and decay, while their spin rates synchronize and rotation axes align
\citep{1980A&A....92..167H}. The non-spherical equilibrium shape of a planet when subjected to its host star's tidal potential depends on the planet's interior structure \citep{1978ppi..book.....Z, 2013ApJ...767..128C}. Direct observations of tidal distortions would place valuable constraints on interior structure models, which are hampered by degeneracies when only bulk masses and radii are available \citep[e.g.][]{2011A&A...528A..18K, 2012A&A...538A.146K}. 

The link between tidal distortion and interior structure is encapsulated in the planet's ``Love numbers'' \citep{1909RSPSA..82...73L, 1911spge.book.....L}. The planet deforms under the combined influence of the external tidal potential from the host star and the centrifugal potential from rotation \citep{2011A&A...528A..41L}. The planet's equilibrium radial displacement $\Delta R$ at its surface is proportional to the perturbing potential $V_p$ according to \citep{2014A&A...570L...5C} 
\begin{equation}
\Delta R = - h_2 V_p/g,
\end{equation}
where $g = G M_p/R_p^2$ is the surface gravity, and $h_2$ is the fluid second-degree radial Love number. The maximum value of $h_2$ is $5/2$ for a homogeneous mass distribution. For a fluid planet, $h_2$ is related to the second-degree potential Love number $k_2$ as $h_2 = 1 +  k_2$ \citep{1939MNRAS..99..451S, 1978ppi..book.....Z}. The value of $k_2$ reflects the degree of central concentration of the planet, with larger values of $k_2$ indicating a more homogeneous mass distribution up to a maximum value of $3/2$ \citep[see, e.g.,][]{ 1978ppi..book.....Z, 2010A&A...523A..26N, 2011A&A...528A..18K, 2018A&A...615A..39K}.


Assuming that the planet's spin axis is aligned and its rotation rate is synchronous, the tidal potential is three times larger than the centrifugal potential, and it dominates the observed shape distortion \citep{2014A&A...570L...5C}, although the tidal and rotational distortions may be difficult to separate observationally \citep{2011A&A...528A..41L}. The equilibrium shape is approximately a triaxial ellipsoid with semi-principal axes $R_1 > R_2 \gtrsim R_3$ \citep{2014A&A...570L...5C}. The three axes are related by $R_1 = R_2(1+3 q)$ and  $R_3 = R_2(1-q)$  where
\begin{equation}
q = \left(\frac{h_2}{2}\right)\left(\frac{M_{\star}}{M_p}\right)\left(\frac{R_2}{a}\right)^3. 
\label{eq: q}
\end{equation}
Tidal deformation increases as the planet-to-star distance shrinks and as the planet's radius increases. Very short-period hot Jupiters are thus the most favorable targets.

One way to measure $k_2$ is to observe the apsidal precession (the gradual change in the orientation of the elliptical orbit) caused by the planet's aspherical shape \citep{2009ApJ...698.1778R}. Possible signatures of apsidal precession were seen for WASP-4\,b \citep[e.g.][]{2019AJ....157..217B, 2020ApJ...893L..29B, 2022AJ....163..281T, 2023A&A...669A.124H}, WASP-18\,b \citep{2019A&A...623A..45C}, and WASP-19\,Ab \citep{2024A&A...684A..78B} using transit timing variations (TTVs) and radial velocity data, but in some cases, the statistical significance is modest and the interpretation of the data is not unique. Recently, JWST photometry helped to constrain apsidal precession of WASP-43\,b \citep{2025A&A...694A.233B}, although the apparent rate is too fast to be attributed solely to the planet's tidal deformation. Because the detectable signatures of apsidal precession are proportional to eccentricity, this method calls for an orbit wide enough to avoid tidal circularization ($P \gtrsim 3$ days for a Jupiter-like planet, as shown in the third bar in Figure \ref{fig: mechanism regimes}) or an orbit that maintains a non-zero eccentricity due to external perturbations. 


Another method to measure the Love number of an exoplanet is to observe its tidal deformation directly in a transit or full-orbit light curve \citep{2009ApJ...698.1778R, 2014A&A...570L...5C, 2019ApJ...878..119H, 2024A&A...682A..15A}.  
The transit light curve of a tidally stretched planet is slightly different from that of a spherical planet of the same cross-sectional area.
Unfortunately, the observable effects are muted because the long axis of the tidal bulge points towards the star, causing the transverse cross section (and the consequent flux diminution of the star) to vary only slightly throughout the transit. For example,
when the change in projected area between mid-transit and quadrature is 10\%, the change in projected area between mid-transit and ingress is only 0.5--1\%. 

\begin{figure*}[th!]
\centering
\includegraphics[width=.87\linewidth]{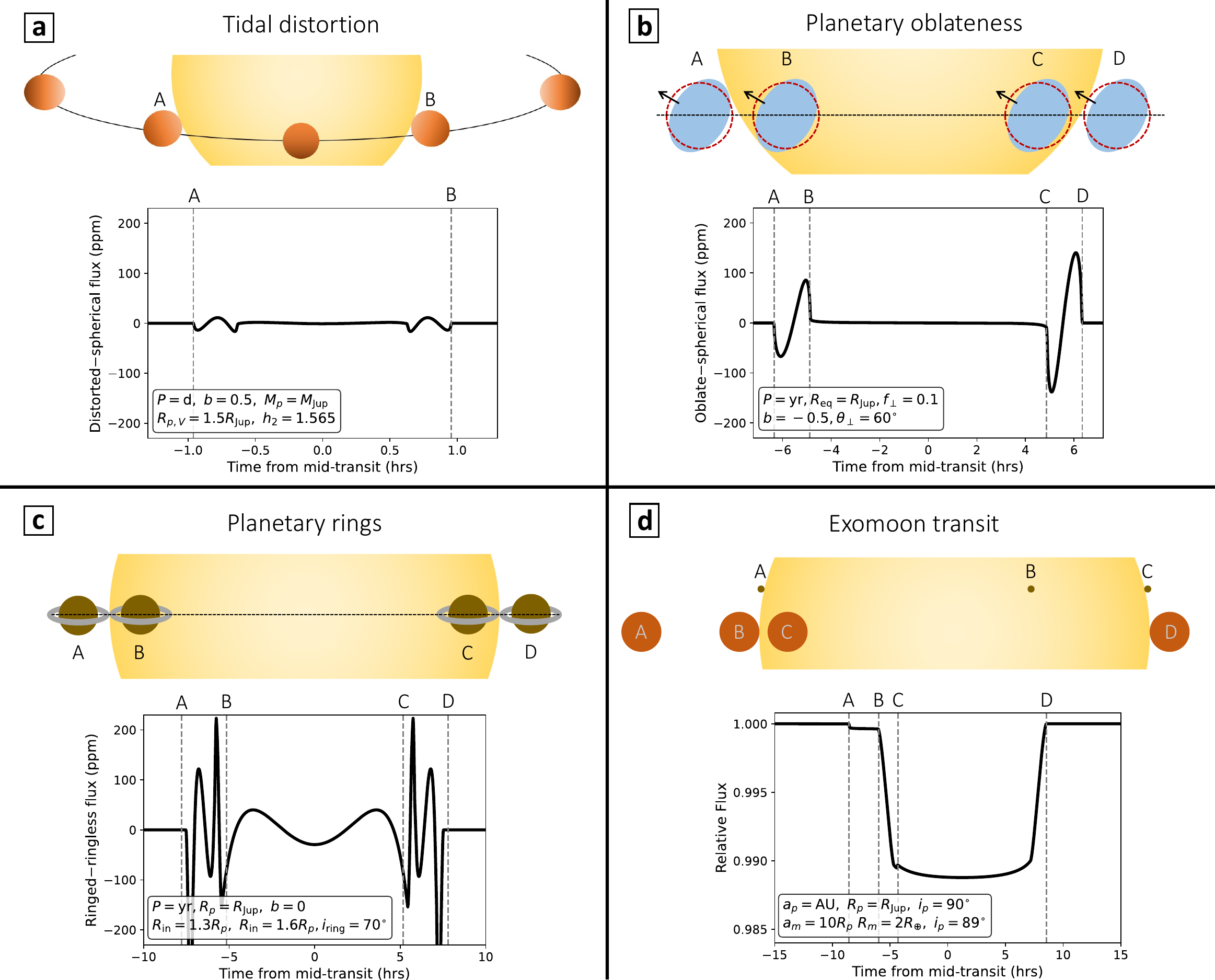}
\caption{\textbf{Illustrations of some phenomena observable in transit light curves.} The host star is assumed to be identical to the Sun; the other relevant parameters are indicated in each panel. The limb darkening parameters are calculated for a wavelength of $3 \mathrm{\mu m}$ using the Exoplanet Characterization Toolkit Limb Darkening Calculator \citep{2021zndo...4556063B}. The illustrations are to scale unless otherwise noted. 
\textbf{(a)} A tidally-distorted short-period planet. Shown are the deviations between the light curve of the distorted planet (calculated with the \texttt{ellc} code \citep{2016A&A...591A.111M}) and the best-fit light curve of a spherical planet.
The planet's size is to scale, including at mid-transit and quadrature, but the orbit is scaled down by a factor of two.
\textbf{(b)} An oblate planet with a tilted spin axis (blue). The red dotted circle has the same projected area as the oblate planet, for comparison.
Shown are the deviations between the light curve of the oblate planet (calculated with the \texttt{JoJo} code \citep{2025AJ....169...79L}) and the best-fit light curve of a spherical planet. The planet's size is $2\times$ larger and the oblateness is $3\times$ larger than the values used to calculate the light curve. 
\textbf{(c)} A planet with a tilted ring system. Shown are the deviations between the light curve of the ringed planet (calculated with the \texttt{pyPplusS} code \citep{2019MNRAS.490.1111R}) and the best-fit light curve of a spherical ringless planet.  
\textbf{(d)} A transit of a Jupiter-sized planet and its super-Earth-sized moon.}
\label{fig:schematic}
\end{figure*}

The main deviations take the form of ``wiggles" in the light curve near ingress and egress (see panel a of Figure \ref{fig:schematic}).
For the case that is illustrated -- a hot Jupiter on a one-day orbit -- the distortion signal varies (from min to max) on a timescale of $\sim10$ min with an amplitude of only $\sim15$ ppm. This is not likely to be detectable with JWST. 
To assess the detectability of tidal distortions more generally, we simulated light curves of distorted, synchronously rotating planets over a range of parameters\footnote{We used the \texttt{ellc} code \citep{2016A&A...591A.111M}, assumed a solar-type star, and sampled planet parameters uniformly in the ranges $ P \in [0.5, 2]\,\mathrm{days}$, $R_{p,V}/R_{\star} \in [1,2]\,R_{\mathrm{Jup}}/R_{\odot}$, $M_p/M_{\star} \in [0.5, 2]\,M_{\mathrm{Jup}}/M_{\odot}$, and $h_2 \in [1.3, 1.7]$.}
and determined for each the maximum amplitude $A_{\rm dis}$ of the deviation between the light curve of the distorted planet and 
the best-fit light curve of a spherical planet.
The results are well described by the scaling law
\begin{equation}
A_{\mathrm{dis}} \approx 18 \ \mathrm{ppm}\left(\frac{a/R_{\star}}{4}\right)^{\!\!-3} \left(\frac{R_{p,V}/R_{\star}}{0.15}\right)^{\!\!5} \left(\frac{M_p/M_{\star}}{0.001}\right)^{\!\!-1} \left(\frac{h_2}{1.5}\right),
\label{eq: distortion signal amplitude}
\end{equation}
where $R_{p,V}$ is the volumetric mean radius.
As expected, the distortion signal is strongest for ultra-short period gas giant planets that are nearly filling their Roche lobes.

For planets where tidal distortion can be photometrically detected, it is possible to constrain the parameter $q$ (equation \ref{eq: q}) and thus the radial Love number $h_2$ \citep{2014A&A...570L...5C}. Several studies have searched for distortion signatures of ultra-hot Jupiters in photometric data from TESS and the CHaracterising ExOplanet Satellite (CHEOPS) \cite{2022A&A...657A..52B, 2024A&A...685A..63A}. Although some detections have been claimed, the statistical significance is modest and the constraints on $h_2$ are weak, motivating further observations. 

As for JWST, inflated gas giant planets on very short-period orbits are the best candidates for detecting tidal distortion. The distortion signal will vary on a timescale of 5--10 min. Given equation \ref{eq: distortion signal amplitude} and the nominal precision of 100 ppm in 1 min cited earlier, we might expect detections in a handful of favorable cases. For example, an upcoming program (GO 5022, PI: Barros) aims to detect the tidal distortion of WASP-103 b using transit and phase curve observations. Another program (GO 9042, PI: Kipping) will reanalyze archival JWST transit data of a sample of hot Jupiters to measure their tidal deformations and constrain their interior structures.

\section*{Tidal inflation}
\label{sec: tidal inflation}


Whenever a planet's tidal distortion varies in time, the planet experiences internal friction that leads to tidal heating. How and where the heat is dissipated are open questions \citep{2014ARA&A..52..171O, 2017ApJ...844...94K}, but it is clear that the extra internal heat can have important ramifications for atmospheric structure and chemistry \citep{2020AJ....160..288F}. Here we explore these effects and describe how JWST could observe them as a means of constraining the tidal properties of exoplanets. While in previous sections our emphasis was Jupiter-sized planets, here we focus more on Neptune-sized planets because their lower surface gravities allow for more readily detected alterations of their atmospheres.

A planet's physical response to tidal dissipation depends on how effectively tidal forces convert orbital or rotational energy into internal heat. As mentioned earlier in the context of tidal orbital decay, the reduced tidal quality factor, $Q'$, quantifies the rate of tidal dissipation. 
With a few exceptions \citep[e.g.][]{2017AJ....153...86M, 2018AJ....155..157P}, $Q'$ is neither measured nor predictable from first principles for exoplanets. Using a novel approach to measure $Q'$ would help constrain planetary interior structures, tidal properties, and orbital evolution \citep{2010A&A...516A..64L, 2019NatAs...3..424M, 2023ApJ...958L..21L}. 

A dramatic physical response to tidal heating is atmospheric inflation \citep{2001ApJ...548..466B, 2019ApJ...886...72M,  2020ApJ...897....7M}. 
For planets with solid cores and $\sim1\%-30\%$ of their mass in H/He-dominated envelopes, tides can inflate the planet's radius up to twice the size it would have in the absence of tidal heating \citep{2019ApJ...886...72M, 2020ApJ...897....7M}. Even modest heating can drive a significant response; radius inflation is expected whenever the tidal luminosity is greater than $\sim10^{-5}$ of the incident stellar luminosity \citep{2019ApJ...886...72M, 2020ApJ...897....7M}. Figure \ref{fig: mechanism regimes} shows an estimate of this regime for nominal system parameters. Unfortunately, measurements of the planet's orbital properties, mass, and radius are not enough to constrain $Q'$. This is because radius inflation also depends on the mass of the planet's gaseous envelope, parameterized with the mass fraction $f_{\mathrm{env}} = M_{\mathrm{env}}/M_p$, which is typically only a few-\% for sub-Neptunes \citep[e.g.][]{2015ApJ...806..183W}. The two quantities $Q'$ and $f_{\mathrm{env}}$  both affect the planet's radius. Therefore, a puffy planet could be explained by either a small $Q'$ or a large $f_{\mathrm{env}}$. In order to constrain $Q'$ or to prove that a given planet is puffy due to tides, it is necessary to use additional information such as atmospheric chemistry.

For this application, unlike the others described in this article, spectroscopy is crucial. The extra internal heating from tides drives vigorous vertical mixing that alters molecular abundances in observable parts of the planet's atmosphere. The observed spectrum depends largely on the atmospheric metallicity, the equilibrium temperature $T_{\mathrm{eq}}$, the intrinsic temperature $T_{\mathrm{int}}$, and the vertical eddy diffusion coefficient $K_{zz}$ \citep{2020AJ....160..288F}. $T_{\mathrm{int}}$ parameterizes the intrinsic flux generated from the planetary interior such that $T_{\mathrm{eff}}^4 = T_{\mathrm{eq}}^4 + T_{\mathrm{int}}^4$ where Jupiter has $T_{\mathrm{int}} \approx 100$ K \citep[e.g.][]{1977Icar...30..305H}. Tidal heating can raise $T_{\mathrm{int}}$ by several hundred kelvin, which drives more vigorous vertical mixing  (larger $K_{\mathrm{zz}}$) and non-equilibrium chemistry \citep{2020AJ....160..288F}. Stronger vertical mixing deepens the ``quench pressure'', beyond which the atmosphere is in chemical equilibrium 
\citep[e.g.][]{2002Icar..155..393L, 2011ApJ...737...15M}. At lower pressures (higher altitudes) than the quench pressure, the molecular species are uniformly mixed and their abundances reflect the equilibrium conditions at the quench pressure rather than the local atmospheric conditions. The observed abundances at the relatively high altitudes and low pressures ($\sim1$ mbar) probed with transit spectroscopy thus serve as a thermometer of the deep interior.

A specific example is the use of methane depletion as an interior thermometer in Neptune or Saturn-like planets. For such planets having equilibrium temperatures of 500--800\,K, CH$_4$ would be the dominant carbon carrier and CO and CO$_2$ would be less prevalent if the atmosphere were in chemical equilibrium. However, strong convection from tidal heating should draw up CH$_4$-poor and CO-rich gas \citep{2020AJ....160..288F}, providing an explanation for prior non-detections of CH$_4$ in the atmospheres of the moderately eccentric Neptune-sized exoplanets GJ 436\,b, GJ 3470\,b, and WASP-107\,b \citep{2010Natur.464.1161S, 2013ApJ...777...34M, 2018ApJ...858L...6K, 2019NatAs...3..813B, 2024Natur.625...51D}. 

This hypothesis was confirmed using JWST observations of the abnormally puffy warm Neptune WASP-107\,b \citep{2024Natur.630..831S, 2024Natur.630..836W}. 
The concentration of methane inferred from a NIRSpec transit spectrum was 1{,}000 times lower than the equilibrium expectation \citep{2024Natur.630..831S}. The planet was found to have a very hot interior ($T_{\mathrm{int}} = 460 \pm 40$ K).
Another study was based on a transit spectrum using HST's Wide Field Camera 3 (WFC3) and JWST's NIRCam and MIRI, which led to the detection of CH$_4$ and other molecules \citep{2024Natur.630..836W}. They also found vigorous vertical mixing and a high $T_{\mathrm{int}}$.
Given the radius inflation driven by the tidally-induced internal heat flux, both studies found that the mass of the solid portion of the planet is more than twice as large as that inferred from models that ignored tides and that the tidal quality factor is $Q' \lesssim 10^4$. These results confirmed that tidal heating is observable through its effects on atmospheric structure and chemistry, boding well for future applications of non-equilibrium chemistry as a thermometer of the deep atmosphere of other planets.\footnote{For instance, programs like GO 2594 (PI: Spake), GO 9095 (PI: Piaulet-Ghorayeb), and GO 9101 (PI: Radica) will apply this method to measure the internal temperatures of more exo-Neptunes.}


\section*{Rotational oblateness and obliquity}

Rotation causes a planet's shape to become oblate, as centrifugal forces lead to bulging at the equator and flattening at the poles. The oblateness is directly related to the planet's rotation rate and density structure. Thus, observing oblateness is a possible means of probing exoplanet rotation and interior structure. Planetary oblateness can be quantified by the flattening parameter 
\begin{equation}
f \equiv \frac{R_{\mathrm{eq}} - R_{\mathrm{pol}}}{R_{\mathrm{eq}}},
\end{equation}
where $R_{\mathrm{eq}}$ and $R_{\mathrm{pol}}$ are the equatorial and polar radii of the planet. Jupiter and Saturn have $f_{\mathrm{Jup}}$ = 0.065 and $f_{\mathrm{Sat}}$ = 0.098.

Assuming the planet is in hydrostatic equilibrium and that its shape is only slightly non-spherical, the Darwin-Radau relation connects flattening with other planet properties \citep{1989Icar...78..102H, 1999ssd..book.....M}:
\begin{equation}
f = \frac{\Omega_p^2}{G}\frac{R_p^3}{M_p}\left[\frac{5}{2}\left(1 - \frac{3}{2}C\right)^2 + \frac{2}{5}\right]^{-1},
\label{eq: f from Darwin-Radau}
\end{equation}
where $\Omega_p = 2\pi/P_{\mathrm{rot}}$ is the planet's rotation rate and $C$ is its polar moment of inertia divided by $M_p R_p^2$. Equation \ref{eq: f from Darwin-Radau} indicates that high oblateness is associated with rapid
rotation and low overall density. Gaseous planets are thought to form with rapid rotation, at rates within 10\%-30\% of their break-up rates $\Omega_{\mathrm{b}} = \sqrt{G M_p/ R_p^3}$ 
\citep{2018AJ....155..178B, 2020ApJ...905...37B}. Close-in planets experience tidal spin-down until their rotation periods are synchronized with their orbital periods. Thus, we expect that the most oblate planets should be young or far enough from their host stars to have avoided tidal spin-down. The timescale for tidal synchronization is \citep{1996Icar..122..166G,
2010A&A...516A..64L}
\begin{equation}
\tau_{\mathrm{sync}} = \left(\frac{\Omega_{p,i}}{n^2}\right)\left(\frac{2 C Q'}{9}\right)\left(\frac{M_p}{M_{\star}}\right)\left(\frac{a}{R_p}\right)^{\!\!3}
\label{eq: tau_sync}
\end{equation}
where $\Omega_{p,i}$ is the planet's initial spin angular velocity. 
For billion-year-old systems, fast spin rates are expected for planets when $\tau_{\mathrm{sync}} \gtrsim \ \mathrm{Gyr}$. Figure \ref{fig: mechanism regimes} shows that this corresponds to $P \gtrsim 30$ days for a Jupiter-like planet around a Sun-like star.

It has long been theorized that oblateness should be detectable using transit photometry, since the transit light curve of an oblate planet differs slightly from that of a spherical planet of the same cross-sectional area \citep{2002ApJ...572..540H,  2002ApJ...574.1004S, 2003ApJ...588..545B, 2010ApJ...709.1219C}. The deviations occur mainly in the ingress and egress phases of the transit, and their detectability depends strongly on the planetary obliquity, $\theta$, the angle between the planet's spin axis and its orbital axis. An oblate planet with a low obliquity is difficult to distinguish from a spherical planet, but a large obliquity can break the usual symmetry of the ingress and egress, a signal that cannot be mimicked by adjusting the parameters of a spherical planet.
The oblateness signal is the difference between the light curve of an oblate planet and the best-fit light curve of a spherical planet. An example is shown in panel b of Figure \ref{fig:schematic} for a tilted, Jupiter-sized planet with a Saturn-like oblateness. 

The light-curve deviations depends not on the true oblateness and obliquity, but rather on their sky projections $f_{\perp}$ and $\theta_{\perp}$.
We estimated the maximum amplitude $A_{\rm obl, max}$ as a function of system parameters by sampling $f_{\perp}$ and $R_p/R_{\star}$ (where $R_p \equiv \sqrt{R_{\mathrm{eq}} R_{\mathrm{pol}}}$), modeling the oblateness signal using the \texttt{JoJo} code \citep{2025AJ....169...79L}, and fitting the resulting signal amplitudes with a generalized power law. This results in 
\begin{equation}
A_{\mathrm{obl}, \mathrm{max}} \approx 150 \ \mathrm{ppm} \left(\frac{R_p/R_{\star}}{0.1}\right)^2\left(\frac{f_{\perp}}{0.1}\right),
\end{equation}
which agrees well with previous work \citep{2014ApJ...796...67Z}. 
This equation gives the maximum amplitude over all possible orientations of the spin axis. More generally, we have 
$A_{\mathrm{obl}} = A_{\mathrm{obl}, \mathrm{max}} g(b, \theta_{\perp})$ where $b$ is the planet's impact parameter and $g(b, \theta_{\perp}) \leq 1$  \cite{2014ApJ...796...67Z}.
The function $g$ is complicated,
but for each value of $b$ there is
a value of $\theta_{\perp}$ for which $g(b, \theta_{\perp}) = 1$. For example, when $b = 0.7$, the optimum value of $\theta_{\perp}$
is $45^{\circ}$. Based on a sample of $\sim 900$ transiting exoplanets with $P > 10$ days, a previous study \cite{2022ApJ...935..178B} found that there should be several dozen exoplanets
with signal amplitudes in the $30-200$ ppm range. Moreover, greater sensitivity can be achieved for planets with high impact parameters \citep{2024arXiv241003449D}.

Signals of this magnitude are theoretically detectable with Kepler photometry, yet to date there have only been upper limits or marginal and unreliable detections \citep{2010ApJ...709.1219C, 2014ApJ...796...67Z, 2017AJ....154..164B}. Multiple factors have hindered confirmation. Kepler photometry has a precision of $\sim100$ ppm for only the brightest targets, $V \lesssim 12$ \citep{2012PASP..124.1279C}, and in most cases the data were averaged into $30$-min samples,
smearing out any oblateness signals during ingress and egress. 


The prospects of detecting planetary oblateness with JWST are promising \citep{2025AJ....169...79L}. For instance, two recent studies used JWST NIRSpec data to constrain the oblateness of the very low density planet Kepler-51\,d \citep{2024ApJ...977L...1L,
2024ApJ...976L..14L}, finding consistent results. The first of these \citep{2024ApJ...977L...1L} placed an upper limit of $f_{\perp} < 0.15$,  
which also resulted in a  constraint on the planet's rotation period, $P_{\mathrm{rot}} \gtrsim 40$ hours. An upcoming program (GO 7188, PI: Acuna) plans to use NIRSpec to measure the oblateness of the long-period gas giant, TOI-199~b. Interest in oblateness detection has been gaining momentum, with additional studies exploring constraints from K2 data \citep{2024arXiv241005408P} and presenting new codes for oblateness studies \citep{2024ascl.soft12025C, 2024arXiv241005408P, 2024arXiv241003449D, 2025AJ....169...79L}. A clear detection of oblateness appears within reach but will require a little bit of luck: a relatively bright and photometrically quiet host star, and a planet with Saturn-like oblateness and a suitably misaligned spin axis. There is still room to increase the sample of long-period planets around bright stars (recall Figure \ref{fig:kepler_vs_allsky}). 

Before moving on, we note that transits are not the only way to detect planetary rotation and obliquities; direct imaging is another viable method for long-period planets \citep[e.g.][]{2020AJ....159..181B} that is being explored with JWST (e.g. GO 4758, PI: Zhou).


\section*{Planetary rings}

Planetary rings might be detected through the anomalies they create in transit light curves. Planetary rings are thin disks composed of dust, ice, and rocky debris. Rings are a probe of the formation histories of planets and their satellites, and they can also reveal properties of the giant planet interiors. The Sun's four giant planets all have rings, although they vary greatly in terms of radial extent, composition, particle size, and other properties \citep[e.g.][]{2018prs..book..517C}. The existence of rings around exoplanets has long been a topic of speculation \citep[e.g.][]{1999CRASB.327..621S, 2001ApJ...552..699B}. A survey of ``exorings'' would help constrain key open questions like how rings form and how long they persist. A recent review was provided by Ref.~\cite{2024arXiv240113293T}.

Exorings are theoretically detectable using a variety of  techniques, including reflected light phase curves and spectroscopy \citep{2004A&A...420.1153A, 2005ApJ...618..973D, 2015A&A...583A..50S}, transit spectroscopy \citep{ 2022ApJ...930...50O}, the Rossiter-McLaughlin effect \citep{2009ApJ...690....1O, 2017MNRAS.472.2713D}, and precise transit light curves \citep{2003ApJ...588..545B, 2009ApJ...690....1O, 2015ApJ...803L..14Z, 2017AJ....153..193A, 2018A&A...609A..21A}. We will focus here on detection with transit light curves. When rings are not viewed edge-on, they increase the depth and duration of the transit and can cause complex light curve deviations concentrated near the ingress and egress phases. 

The detectability of exorings with transit photometry is more complicated to quantify than the detectability of planetary oblateness because rings have more unknown parameters, such as the ring size, opacity, and inclination. 
Only massive, Saturn-like rings are plausibly detectable with JWST. Although Saturn's rings are primarily composed of water ice \citep{2010Sci...327.1470C, 2013pss3.book..309T}, rocky rings might exist around planets interior to the ice line \citep{2011ApJ...734..117S}. Saturn-like rings would produce a signal reaching up to $\sim300-500$ ppm, when measured as the difference between a ringed planet's transit light curve and the light curve of the best-fit ringless planet \citep[e.g.][]{2003ApJ...588..545B, 2009ApJ...690....1O}. Detecting exorings requires $\sim100$ ppm photometric precision on $\sim10$ minute timescales \citep{2018A&A...609A..21A, 2020AJ....159..131P}. An example of a photometric signal of planetary rings is shown in panel c of Figure \ref{fig:schematic} for a Jupiter-sized planet with tilted, opaque rings extending from $1.3 - 1.6 \ R_p$.

Planetary rings are more likely to be detectable around planets with longer periods. One reason is that the planet needs to avoid tidal spin-orbit alignment, which would otherwise cause the rings to be in the nearly invisible edge-on configuration. At short orbital periods, planetary obliquities are lowered by tides unless they are trapped in a secular spin-orbit resonance \citep{2019NatAs...3..424M, 2022MNRAS.509.3301S}. Tidal realignment of a planetary obliquity occurs on a similar timescale as tidal synchronization, so the required orbital period range for maintaining a large obliquity is similar to that for oblateness ($P \gtrsim 30$ days for a Jupiter-like planet as shown in Figure \ref{fig: mechanism regimes}). Another factor favoring longer orbital periods is that rings are only stable in a tilted planet's equatorial plane if the planet is sufficiently oblate, which in turn requires a planet that has avoided tidal spin-down \citep{2009AJ....137.3706T, 2020AJ....159..131P}.

Exoplanets with transit-derived radii much larger than expected are good candidates to search for rings. It has been proposed that some of the ``super-puff'' planets --- those with bulk densities that appear to be below about $0.1 \ \mathrm{g} \ \mathrm{cm}^{-3}$ --- could be planets of ordinary density surrounded by rings \citep[e.g.][]{2020A&A...635L...8A, 2020AJ....159..131P, 2022ApJ...930...50O, 2023A&A...675A.174S}. The planet HIP 41378 f is the most studied candidate in this regard \citep{2020AJ....159..131P, 2020A&A...635L...8A, 2023A&A...675A.174S, 2024arXiv241000641L}. HIP 41378 f has a period of 542 days and a measured mass and radii of $12 \pm 3 \ M_{\oplus}$ and $9.2 \pm 0.1 \ R_{\oplus}$, yielding a density of $0.09 \ \mathrm{g} \ \mathrm{cm}^{-3}$ \citep{2019arXiv191107355S}. However, models for a ringed planet were assessed using data from the K2 Mission \citep{2020A&A...635L...8A}, and there was no conclusive evidence for characteristic ring signatures near the ingress or egress of the transit.

Systematic searches for exorings have been conducted in several studies, especially using Kepler data \citep[e.g.][]{2015ApJ...814...81H, 2017AJ....153..193A}, with no confirmed detections so far. The main limitation has been sufficient photometric precision over short timescales and the degeneracy of possible ring solutions with other models. JWST can improve on the precision. The photometric signal can reach several 100 ppm for tilted, extensive rings as pictured in Figure \ref{fig:schematic} and would be readily detectable with JWST with a nominal precision of 100 ppm in 1 min. As for the degeneracy, the best approach would be to combine photometric and spectral methods like transit spectroscopy \citep[e.g.][]{2022ApJ...940L..30O}.





\section*{Moons}


The planets in the Solar System host a diverse array of moons, tallying 421 to date.\footnote{Retrieved March 22, 2025 from https://ssd.jpl.nasa.gov/} Most of these moons are smaller than 100 km, but at least 19 of them are massive enough for gravity to cause the moon to attain a nearly spherical shape.
It is natural to expect moons to exist around exoplanets, and indeed, ``exomoons'' are a perennial topic in the exoplanet literature \citep{1999A&AS..134..553S, 2006A&A...450..395S, 2007A&A...470..727S, 2007A&A...464.1133C, 2009MNRAS.392..181K, 2009MNRAS.396.1797K}. However, no confirmed detections exist yet. For a review of exomoons, see Ref.~\cite{2024arXiv240113293T}. Here, we focus on theoretical expectations for moon system properties, photometric detection strategies, and promising targets. 

Which types of exoplanets are most likely to host moons? In general, moons are most likely to be found around gas giants planets with $P\gtrsim 300$ days \citep[e.g.][]{ 2015A&A...578A..19H,
2016ApJ...817...18S, 2022NatAs...6..367K}. This is partially because the only stable orbits for moons are well within the planet's Hill radius,
\begin{equation}
R_H \approx a \left(\frac{M_p}{3M_{\star}}\right)^{\frac{1}{3}} \approx 143 \ \mathrm{R_{\mathrm{Jup}}} \left(\frac{a}{\mathrm{AU}}\right)\left(\frac{M_p}{M_{\mathrm{Jup}}}\right)^{\frac{1}{3}}\left(\frac{M_{\star}}{M_{\odot}}\right)^{-\frac{1}{3}}
\end{equation}
where this equation assumes $M_p \ll M_{\star}$. Moons within the Solar System are well within this limit, with median $a_m/R_H \approx 0.008$ and maximum $a_m/R_H \approx 0.25$ for the 19 major moons. Prograde moons are only stable for $a_m/R_H \lesssim 0.4-0.5$ \citep{2006MNRAS.373.1227D, 2020AJ....159..260R, 2024MNRAS.527.4371K}. Since $R_H\propto a$, moons seem unlikely to exist around very close-in planets.

Long-term orbital stability is also affected by tidal evolution of the moon's orbit, which is primarily driven by the tides the moon raises in the planet. Moons with orbital periods smaller than the planet's rotation period ($P_m < P_{\mathrm{rot}}$) spiral inwards until they cross the Roche limit and are disrupted, while those with $P_m > P_{\mathrm{rot}}$ spiral outwards and may escape from the planet \citep{2002ApJ...575.1087B, 2006MNRAS.373.1227D, 2012ApJ...754...51S, 2017MNRAS.471.3019A, 2021PASP..133i4401D}, unless escape can be avoided through dissipation inside the moon \citep{2024MNRAS.527.4371K}. As a result, \cite{2021PASP..133i4401D} found that most planets with $P < 10$\,days cannot hold long-term stable moons, and the survival fraction increases from $\sim0\%$ at $P = 10$ days to $\sim70\%$ at $P = 300$ days. This is depicted in Figure \ref{fig: mechanism regimes}, though we note that this is an average statement and subtleties exist. 

What is the highest exomoon mass we can reasonably expect? Those that formed via core accretion in primordial circumplanetary disks are theoretically expected to form up to mass ratios of $M_m/M_p \sim 10^{-4}$ \citep{2006Natur.441..834C,
2020ApJ...894..143B, 2021MNRAS.504.5455C}, as seen in the satellites of the Solar System's outer planets. Moons formed via giant impacts or gravitational capture could theoretically form with larger mass fractions under the right conditions, although there is substantial theoretical uncertainty  \citep[e.g.][]{2014ApJ...790...92O, 2018ApJ...869L..27H, 2019SciA....5.8665H, 2020MNRAS.492.5089M, 2022NatCo..13..568N}. 

The transit method offers the best chance for a secure detection and subsequent characterization of an exomoon. The moon could be directly observed in transit before, after, or during the planetary transit. An example of what this might look like is shown in panel d of Figure \ref{fig:schematic}. Detection in phase-folded data has also been proposed \citep{2014ApJ...787...14H}. Galilean-sized moons will likely only be detectable around K or M dwarfs with this method. For instance, Jupiter's largest moon Ganymede, which has a mass $0.025 \ M_{\oplus} \approx 8 \times 10^{-5} \ M_{\mathrm{Jup}}$ and radius $0.4 \ R_{\oplus}$ would produce a transit depth of only 14 ppm around a Sun-like star but 90 ppm around a $0.4 \ M_{\odot}$ star.

Additionally, moons could be detected indirectly by the TTVs and transit duration variations (TDVs) they induce on the planets \citep{2007A&A...470..727S, 2009MNRAS.392..181K, 2009MNRAS.396.1797K, 2016A&A...591A..67H}. The TTV amplitude can be detectable for a sufficiently massive and widely-separated moon, scaling as \citep{2021MNRAS.500.1851K}
\begin{equation}
\mathrm{A_{TTV}} = 34 \ \mathrm{s}\left(\frac{a_m}{R_H}\right)\left(\frac{M_p/M_{\star}}{M_{\mathrm{Jup}}/M_{\odot}}\right)^{\frac{1}{3}}\left(\frac{M_m/M_p}{10^{-4}}\right)\left(\frac{P}{\mathrm{yr}}\right),
\end{equation}
where here we assume circular orbits.
Moreover, the TTV periods $P_{\mathrm{TTV}}$ tend to be very short and frequently fall in a regime in which TTVs from planet-planet interactions are rarely seen \citep{2021MNRAS.500.1851K, 2023MNRAS.518.3482K}. The TDV amplitude (again for circular orbits) is 
\begin{equation}
\mathrm{A_{TDV}} = 0.60 \ \mathrm{s}\left(\frac{a_m}{R_H}\right)^{\!\!-\frac{1}{2}}\left(\frac{M_p/M_{\star}}{M_{\mathrm{Jup}}/M_{\odot}}\right)^{\!\!\frac{1}{3}}\left(\frac{M_m/M_p}{10^{-4}}\right)\left(\frac{t_T}{14 \ \mathrm{hr}}\right).
\end{equation}
The TDV signal is more challenging to detect, but would be helpful
for confirmation if they could be observed in concert with TTVs.

There are no confirmed detections of exomoons to-date, although there have been several candidates proposed over the years, including some from microlensing and transit observations. We refer to \cite{2024arXiv240113293T} for a full discussion but highlight the photometrically-identified candidates around the long-period gas giant planets Kepler-1625 b \citep{2018SciA....4.1784T} and Kepler-1708 b \citep{2022NatAs...6..367K}. Both proposed moon candidates are large, roughly Neptune-sized bodies. Future observations are still necessary to validate or reject them.  

Several JWST programs are slated to search for exomoons using the transit method. One program (GO 6491, PI: Cassese) will search for moons around the long-period Jupiter-like planet Kepler-167 e by observing it for 40 hours spanning a transit. Based on the forecasted precision, a Ganymede-sized moon is likely to be detectable. Another program (GO 6193, PI: Pass) will target the Earth-sized planets TOI-700 d and TOI-700 e with sufficient sensitivity to detect analogs of the Moon. These two planets lie within the habitable zone of their M dwarf host, giving this system extra appeal.

\acknow{We gratefully acknowledge the three anonymous reviewers for their feedback and Eric Agol, Ben Cassese, Shashank Dholakia, Shishir Dholakia, Caleb Lammers, Tiger Lu, and David Kipping for comments on an early draft. We thank the following people for their publicly available codes used in this work: Pierre Maxted for \texttt{ellc}, Quanyi Liu and Wei Zhu for \texttt{JoJo}, and Edan Rein and Aviv Ofir for \texttt{pyPplusS}.}

\showacknow 

\newpage
\bibliography{main}

\end{document}